\documentclass[12pt]{article}

\usepackage{graphicx}

\usepackage{amsmath,amsfonts}
\usepackage{bm}
\usepackage{multirow}

\title{Boundary element dynamical energy analysis: a versatile method for solving two or three dimensional wave problems in the high frequency limit}

\author{D.J. Chappell\footnote{david.chappell@nottingham.ac.uk}, G. Tanner\footnote{gregor.tanner@nottingham.ac.uk} and S. Giani\footnote{stefano.giani@nottingham.ac.uk}\\
School of Mathematical Sciences, University of Nottingham,\\
University Park, Nottingham NG7 2RD, UK}
\date{}
\begin{document}

\maketitle

\begin{abstract}
Dynamical energy analysis was recently introduced as a new method for determining the distribution of mechanical and acoustic wave energy in complex built up structures. The technique interpolates between standard statistical energy analysis and full ray tracing, containing both of these methods as limiting cases. As such the applicability of the method is wide ranging and additionally includes the numerical modelling of problems in optics and more generally of linear wave problems in electromagnetics. In this work we consider a new approach to the method with enhanced versatility, enabling three-dimensional problems to be handled in a straightforward manner. The main challenge is the high dimensionality of the problem: we determine the wave energy density both as a function of the spatial coordinate and momentum (or direction) space. The momentum variables are expressed in separable (polar) coordinates facilitating the use of products of univariate basis expansions. However this is not the case for the spatial argument and so we propose to make use of automated mesh generating routines to both localise the approximation, allowing quadrature costs to be kept moderate, and give versatility in the code for different geometric configurations.
\end{abstract}

\section{Introduction}

Predicting the wave energy distribution of the vibro-acoustic response of a complex mechanical system is a challenging task, especially in the mid-to-high frequency regime. Standard numerical tools such as finite element methods become inefficient, and ray or thermodynamic approaches are often employed to model the wave energy flow through the structure. Popular methods are \textit{Statistical Energy Analysis} (SEA) \cite{RL69,RL95, RC96}, in which the mean energy flow between subsystems is assumed to be proportional to the energy gradient, and the \textit{ray tracing technique}, in which the wave intensity distribution is determined by summing over contributions of a potentially large number of ray paths \cite{AG89, HK00, VC01}.

SEA is in fact a low resolution ray tracing method \cite{SK01,GT09} leading to small numerical models compared to ray tracing. This efficiency saving comes at a price, however: SEA has no spatial resolution of the energy distribution within subsystems and becomes unreliable whenever long range correlations in the ray dynamics are present. The recently developed \textit{Dynamical Energy Analysis} (DEA) \cite{GT09, DC11} provides a tool which interpolates between SEA and a full ray tracing analysis and can overcome some of the problems mentioned above at a relatively small computational overhead. DEA thus enhances the range of applicability of standard SEA and gives bounds on the range of applicability of SEA. Related methods have been discussed previously in the context of wave chaos \cite{GT07} and structural dynamics \cite{KH94}. In particular Langley's \textit{Wave Intensity Analysis} (WIA) \cite{RL92, RL94} and Le Bot's thermodynamical high frequency boundary element method \cite{AL98, AL02, AL06} include details of the underlying ray dynamics. The approach employed here differs from these methods by considering multiple reflections in terms of linear operators. Representing these operators in terms of basis function expansions then leads to SEA-type equations.

In this work we develop a new approach to DEA suitable for modelling three-dimensional problems. The present DEA methods rely on the fact that one can easily parametrise the boundary of the region being modelled, and then apply an orthonormal basis approximation over the resulting boundary phase space coordinate system. In two dimensions this is simple as the boundary may be parametrised along its arc-length and the associated momentum (or direction) coordinate taken tangential to the boundary. The basis can be any suitable (scaled) univariate basis in both position and momentum, such as a Fourier basis \cite{GT09} or Chebyshev polynomials \cite{DC11}. Defining a suitable parametrisation for the spatial coordinate in three-dimensions becomes much more difficult. In momentum space spherical polar coordinates may be employed and so these problems do not arise.

In order to develop a flexible code we employ automated mesh generating routines to provide a widely applicable parametrisation of the boundary surface for general three-dimensional structures via triangulation. The precision of the spatial approximation may then be improved by refining the mesh, avoiding the issue of finding a suitable basis. One avenue for potential future study stems from the fact that it is possible to define an orthogonal basis on a general triangle which reduces to Legendre polynomials along one edge of the domain triangle \cite{RF03}. However, in this work we restrict to a piecewise constant approximation on each element of the mesh for reasons of both simplicity and to keep the associated quadrature costs moderate for the three dimensional case.

For the choice of momentum basis we may take a product univariate basis as mentioned above. It is preferable if this basis is orthogonal with respect to the standard $L^2$ inner product for consistency with both the piecewise constant spatial approximation, and the SEA limit when the lowest order momentum basis is applied and continuity is enforced across the mesh. The main choices are either a Fourier basis or Legendre polynomials. In this work we choose Legendre polynomials due to better convergence properties in the absence of periodic boundary conditions \cite{JB00} and for consistency with the approach in \cite{RF03} should we wish to include a spatial basis in future work.

The remainder of the paper is structured as follows. In Section \ref{sec:green}, the ray tracing approximation is discussed and related to the Green function using short wavelength asymptotics. In Section \ref{sec:PF}, the concept of phase-space operators is introduced in order to represent the propagation of ray densities in terms of boundary integrals. The discretization of the method using spatial meshing procedures and basis function approximations in direction space is then detailed. Decomposition of the method for problems with multiple subsystems is then discussed along with links between the method and SEA. In Section \ref{sec:numerics} the application of boundary element DEA to two-dimensional examples is discussed and verified against previous work. Finally some three-dimensional examples are considered.

\section{\label{sec:green} Wave equations and asymptotics}

It is assumed that the system as a whole is characterized by a linear wave equation describing the overall wave dynamics including damping and radiation in a finite domain $\Omega\subset\mathbb{R}^{d}$, $d=2$ or 3. In this work only stationary problems with continuous, monochromatic energy sources are considered. We split the system into $N_{\Omega}$ subsystems and consider the scalar wave equation for acoustic pressure waves in each homogeneous sub-domain $\Omega_{i}$, with local wave velocity $c_{i}$, $i=1,...,N_{\Omega}$ and $\Omega=\bigcup_{i=1}^{N_{\Omega}}\Omega_{i}$. Extensions to more complicated systems with different wave operators in different parts of the system can be treated with the same techniques as long as the underlying wave equations are linear, see the discussion in Ref. \cite{GT09}.

The general problem of determining the response of a system to external forcing with angular frequency $\omega$ at a source point $r_{0}\in\Omega_{0}$ can then be reduced to solving
\begin{equation}\label{wave-eq}
(k_{i}^{2}-\hat{H})G(r,r_{0};\omega)=-F_{0}\:\delta(r-r_{0}),\hspace{5mm}i=1,...,N_{\Omega},
\end{equation}
with $\hat{H} = -\Delta$.  The Green function $G$ represents an acoustic pressure wave and $F_{0}$ is a forcing amplitude term with units $kg\:s^{-2}$ which is just set to unity for simplicity. The solution point is denoted $r\in\Omega_{i}$ and $\delta$ is the Dirac delta distribution. Furthermore, $k_i = \omega/c_i + i \mu_i/2$ is a complex valued wavenumber,
where the imaginary part represents a subsystem dependent damping coefficient
$\mu_{i}$. Throughout this work we take $i=\sqrt{-1}$ unless used as a
subscript, in which case it is an index over the number of subsystems.
The wave energy density induced by the source is then given as
\begin{equation}\label{en-eq1}
\varepsilon(r, r_{0}; \omega)=\frac{|G(r,r_{0};\omega)|^{2}}{\varrho_{i}c_{i}^{2}},
\end{equation}
for $r \in \Omega_i$ where $\varrho_{i}$ is the density of the medium in $\Omega_{i}$.
The linear wave operator $\hat{H}$ can naturally be associated with the underlying ray dynamics via the Eikonal approximation; for more detailed derivation, see Ref. \cite{GT09, OR07, PC09}. Using small wavelength asymptotics, the Green function in equation (\ref{wave-eq}) may be written as a sum over \textit{all} classical rays from $r_{0}$ to $r$ for fixed kinetic energy of the hypothetical ray particle. One obtains \cite{PC09, MG90}
\begin{equation}\label{Grn-atc}
G(r,r_{0};\omega)=\frac{\pi}{(2\pi i)^{(d+1)/2}}\sum_{j:r_{0}\rightarrow r}A_{j}e^{i(k_{i}L_{j}-i\nu_{j}\pi/2)},
\end{equation}
where $L_{j}$ is the length of the ray trajectory between $r_{0}$ and $r$ including possible reflections on boundaries. The amplitudes $A_{j}$ may be written as a product of three terms as in Ref. \cite{GT09} due to damping, mode conversion and reflection/transmission coefficients, and geometrical factors. The phase index $\nu_{j}$ contains contributions from the reflection/transmission coefficients at interfaces between subsystems and from caustics along the ray path.

Analogous representations to (\ref{Grn-atc}) have been considered in detail in quantum mechanics \cite{MG90} and are also valid for general wave equations in elasticity, see Ref. \cite{GT07} for an overview. In the latter case $G$ becomes matrix valued. Note that the summation in equation (\ref{Grn-atc}) is typically over infinitely many terms, where the number of contributing rays increases (in general) exponentially with the length of the trajectories included. This gives rise to convergence issues, especially in the case of low or no damping \cite{GT07}.

The wave energy density (\ref{en-eq1}) can now be expressed as a double sum over classical trajectories and hence
\begin{equation}\label{en-eq2}
\begin{array}{ll}
{\varepsilon(r, r_{0}; \omega)} & {\displaystyle =C\sum_{j,\: j': r_{0}\rightarrow r}A_{j}A_{j'}e^{ik_{i}[L_{j}-L_{j'}]-i[\nu_{j}-\nu_{j'}]\pi/2}}\vspace{2mm}\\
{} & {=C\, [\rho(r,r_{0};\omega)+\textrm{off-diagonal\:terms}],}
\end{array}
\end{equation}
with $C=\pi^{2}/(\varrho_{i}c_{i}^{2}(2\pi)^{(d+1)})$. The dominant contributions to the double sum arise from terms in which the phases cancel exactly; one thus splits the calculation into a diagonal part
\begin{equation}\label{rho-def}
\rho(r,r_{0};\omega)=\sum_{j:r_{0}\rightarrow r}|A_{j}|^{2}
\end{equation}
where $j=j'$, and an off-diagonal part. The diagonal contribution gives a smooth background signal and the off-diagonal terms give rise to fluctuations on the scale of the wavelength. The phases related to different trajectories are (largely) uncorrelated and the resulting net contributions to the off-diagonal part are in general small compared to the smooth part, especially when averaging over frequency intervals of a few wavenumbers.

It has been shown in Ref. \cite{GT09} that calculating the smooth diagonal part (\ref{rho-def}) is equivalent to a ray tracing treatment. That is, the smooth part of the energy density can be described in terms of the flow of fictitious non-interacting particles emerging from the source point $r_{0}$ uniformly in all directions and propagating along ray trajectories. This makes it possible to relate wave energy transport with classical flow equations and thus thermodynamical concepts, which are at the heart of an SEA treatment. In DEA the classical flow is expressed in terms of linear \textit{phase space operators} as detailed in the next section.

\section{\label{sec:PF} Boundary Integral Formulation}

\subsection{Phase Space Boundary Integral Formulation}
Following a purely kinetic viewpoint based on the interpretation that rays are trajectories of particles following Hamiltonian dynamics as detailed in Section 2 of Ref. \cite{OR07}, the time dependence of a density of ray trajectories (or particles) $\tilde{\rho}$ is known to satisfy the Liouville equation
\begin{equation}\label{liouville}
\frac{\partial\tilde{\rho}}{\partial\tau}(X,\tau)+\frac{d X}{d\tau}\cdot\nabla_{X}(\tilde{\rho}(X,\tau))=0,
\end{equation}
where $X=(r,p)$ denotes the phase space coordinate with position $r$ and momentum $p$.
The propagator for the Liouville equation is given by $K^{\tau}(X,Y)=\delta(X-\varphi^{\tau}(Y))$ and is the kernel of a linear phase space operator known as a Perron-Frobenius operator in dynamical systems theory \cite{PC09, JD96}. The phase space flow $\varphi^{\tau}(Y)$ gives the position of the particle after time $\tau$ starting at $Y=(r',p')$ when $\tau=0$. Hence we may write
\begin{equation}\label{fpo1}
\tilde{\rho}(X,\tau)=\int_{\mathbb{P}}K^{\tau}(X,Y)\tilde{\rho}_{0}(Y)dY,
\end{equation}
where $\tilde{\rho}_{0}$ denotes the initial ray density at time $\tau=0$. The domain of integration is over the whole of phase space $\mathbb{P}=\Omega\times \mathbb{R}^d$, where the integration over $\mathbb{R}^{d}$ takes care of the momentum coordinates $p$. Note that the flow satisfies the Hamilton equations of motion given by the system of ordinary differential equations (ODEs)
\begin{equation}\label{rayeq}
\frac{d X}{d\tau}=\left(\begin{array}{rr}0 & 1 \\
-1 & 0\end{array}\right)\nabla_{X} H,
\end{equation}
where $H=|p|^2$ is the Hamilton function for the wave operator $\hat{H}$ in (\ref{wave-eq}). That is to say, substituting $\varphi^{\tau}(Y)$ for $X(\tau)$ in (\ref{rayeq}) satisfies the system of ODEs with $X(0) = Y$.

Consider a source localized at a point $r_{0}$ emitting waves continuously at a fixed angular frequency $\omega$. Standard ray tracing techniques estimate the wave energy at a receiver point $r$ by determining the density of rays starting at $r_{0}$ and reaching $r$ after some unspecified time. This may be written in the form
\begin{equation}\label{fpo2}
\rho(r,r_{0},\omega)=\small{\int_{0}^{\infty}\int_{\mathbb{R}^d}\int_{\mathbb{P}}} w(Y,\tau)K^{\tau}(X,Y)\rho_{0}(Y,\omega)dYdp\:d\tau,
\end{equation}
with initial density $\rho_{0}(Y,\omega)=\delta(r'-r_{0})\delta(k_0^{2}-H(Y))$, where $k_0$ is the wave number at the source point as defined in Eqn.\ (\ref{wave-eq}). It can be shown that equation (\ref{fpo2}) is equivalent to the diagonal approximation (\ref{rho-def}) \cite{GT09}. A weight function $w$ is included to incorporate damping and reflection/transmission coefficients. It is assumed that $w$ is multiplicative, that is, ($w(\varphi^{\tau_{1}}(X),\tau_{2})w(X,\tau_{1})=w(X,\tau_{1}+\tau_{2})$), which holds for standard absorbtion mechanism and reflection processes \cite{PC09}.

In order to solve the stationary flow problem we may rewrite equation (\ref{fpo2}) in boundary integral form using a boundary mapping technique. For the time being let us consider a problem with a single (sub-)system $\Omega=\Omega_{1}$ with boundary $\Gamma$. The boundary mapping procedure involves first mapping the ray density emanating continuously from the source onto the boundary $\Gamma$. The resulting boundary layer density $\rho_{\Gamma}^{(0)}$ is equivalent to a source density on the boundary producing the same ray field in the interior as the original source field after one reflection. Secondly, densities on the boundary are mapped back onto the boundary by a boundary operator $\mathcal{B}$ with kernel $K_{\Gamma}(X^{s},Y^{s};\omega)=w(Y^{s})\delta(X^{s}-\phi^{\omega}(Y^{s}))$, where $X^{s}=(s,p_{s})$ represents the coordinates on the boundary. That is, $s$ parameterizes $\Gamma$ and $p_{s}\in B^{d-1}_{|p|}$ denotes the momentum component tangential to $\Gamma$ at $s$ for fixed $H(X) = |p|^2$, where $B^{d-1}_{|p|}$ is an open ball in $\mathbb{R}^{d-1}$ of radius $|p|$ and centre $s$. Likewise, $Y^{s}=(s',p_{s}')$ and $\phi^{\omega}$ is the invertible boundary map. Note that convexity is assumed to ensure $\phi^{\omega}$ is well defined; non-convex regions can be handled by introducing a cut-off function in the shadow zone as in Ref. \cite{AL02} or by subdividing the regions further.

The stationary density on the boundary induced by the initial boundary distribution $\rho_{\Gamma}^{0}(X^{s},\omega)$ can then be obtained using
\begin{equation}\label{Neum-ser}
\rho_{\Gamma}(\omega)=\sum_{n=0}^{\infty}\mathcal{B}^{n}(\omega)\rho_{\Gamma}^{0}(\omega)=(I-\mathcal{B}(\omega))^{-1}\rho_{\Gamma}^{0}(\omega),
\end{equation}
where $\mathcal{B}^{n}$ contains trajectories undergoing $n$ reflections at the boundary. The resulting density distribution on the boundary $\rho_{\Gamma}(X^{s},\omega)$ can then be mapped back into the interior region. One obtains the density (\ref{fpo2}) after projecting down onto coordinate space.

\subsection{Discretization and Basis Representation}

The long term dynamics are thus contained in the operator $(I-\mathcal{B})^{-1}$ and standard properties of Perron-Frobenius operators ensure that the sum over $n$ in equation (\ref{Neum-ser}) converges for non-vanishing dissipation, or in open systems. In order to evaluate $(I-\mathcal{B})^{-1}$ a finite dimensional approximation of the operator $\mathcal{B}$ must be constructed. In Ref. \cite{GT09, DC11} basis expansions have been applied in both position and momentum coordinates, which is straightforward to implement using univariate expansions in each argument for $\Omega\subset\mathbb{R}^{2}$. However, it is not straightforward to construct a general orthogonal basis with independent spatial arguments when $\Omega\subset\mathbb{R}^{3}$. For this reason we employ a boundary element triangulation of $\Gamma$, with a zero order basis approximation on each element for any $L^{2}-$orthonormal basis, which essentially results in a scaled piecewise constant boundary element approximation. This type of approximation is also often referred to as Ulam's method \cite{JD96}, although here such an approximation would be performed in full phase space, rather than just in its spatial component.

For the approximation in the momentum argument we choose a basis orthogonal in $L^{2}$ for consistency with the spatial approximation. We choose a Legendre polynomial basis for this purpose due to good convergence properties without requiring periodic boundary conditions \cite{JB00}. Note that for $\Omega\subset\mathbb{R}^{2}$, then $p_{s}\in(-|k_{i}|,|k_{i}|)$, and for $\Omega\subset\mathbb{R}^{3}$ then $p_{s}\in[0,|k_{i}|)\times[-\pi,\pi)$. Denote by $\tilde{p}_{s}$ a re-scaling of $p_{s}$ to $(-1,1)$ or $[-1,1)\times[-1,1)$ for the two and three-dimensional cases, respectively. Let us also denote
\begin{equation}\label{2DLeg}
\tilde{P}_{m}(p_s)=\frac{1}{\sqrt{k_{i}}}P_{m}(\tilde{p}_{s})
\end{equation}
for $\Omega\subset\mathbb{R}^{2}$, where $P_{m}$ is the Legendre polynomial of order $m$. For $\Omega\subset\mathbb{R}^{3}$, $m=(m_1,m_2)$ is a multi-index of non-negative integers. Let us write $p_s=(p_k,p_\theta)$ and likewise $\tilde{p}_s=(\tilde{p}_k,\tilde{p}_\theta)$. Denote \begin{equation}\label{3DLeg}
\tilde{P}_{m}(p_s)=\frac{2}{k_{i}\sqrt{\pi}}P_{m_{1}}(\tilde{p}_k)P_{m_{2}}(\tilde{p}_\theta).
\end{equation}
Explicitly the overall approximation is then of the form
\begin{equation}
\rho_{\Gamma}(X^{s},\omega)\approx\sum_{l=1}^{n}\sum_{m=0}^{N} \rho_{l,m} b_{l}(s) \tilde{P}_{m}(p_{s}),
\end{equation}
where $N$ is the order of the basis expansion, $n$ is the number of elements and $b_{l}$ denotes the scaled (for orthonormality in an $L^2$ inner product) piecewise constant boundary element basis function $b_l(s)=2^{(1-d)/2}/\sqrt{A_{l}}$ for $s$ in element $l$, and zero elsewhere. Here $A_{l}$ is the surface area of element $l$ in the three-dimensional case and the length of element $l$ in the two dimensional case for $l=1,..,n$. Note that in the three-dimensional case the sum over $m$ is a double sum over the multi-index. An analogous approximation is also made for $\rho^{0}_{\Gamma}$ and the values of $\rho_{l,m}$ are to be determined by solving the resulting linear system.

The matrix approximation $B$ of $\mathcal{B}(\omega)$ for the case $\Omega\subset\mathbb{R}^{2}$ is therefore given by
\begin{equation}\label{op_approx}\begin{array}{l}
{B_{m+1+N(l-1),\beta+1+N(\alpha-1)}=}\vspace{2mm}\\
{\displaystyle {\small\frac{(2m+1)}{4}\int_{\partial\mathbb{P}}\int_{\partial\mathbb{P}}} \tilde{P}_{m}(p_{s})b_{l}(s)K_{\Gamma}(X^{s},Y^{s};\omega)\tilde{P}_{\beta}(p'_{s})b_{\alpha}(s')dY^{s}dX^{s}}=\vspace{2mm}\\
{\displaystyle\frac{(2m+1)}{4}\int_{\partial\mathbb{P}} w(Y^{s})\tilde{P}_{m}(\phi^{\omega}_{p}(Y^{s}))b_{l}(\phi^{\omega}_{s}(Y^{s}))\tilde{P}_{\beta}(p'_{s})b_{\alpha}(s')dY^{s}.}
\end{array}
\end{equation}
Here we write $\phi^{\omega}=(\phi^{\omega}_{s},\phi^{\omega}_{p})$, to denote the splitting of the position and momentum parts of the boundary map. Also $\partial\mathbb{P}=\Gamma\times B^{d-1}_{|p|}$ is the phase space on the boundary at fixed ``energy" $H(X) = |p|^2$. The only changes for the three-dimensional case are that the indexing is slightly more complicated due to $m$ and $\beta$ becoming multi-indices (hence the pre-factor becomes $(2m_{1}+1)(2m_{2}+1)/16$) and the definition of $\tilde{P}_{m}$ changes from (\ref{2DLeg}) to (\ref{3DLeg}). Obtaining the boundary map is not always straightforward, particularly for general three-dimensional geometries, and hence we write the operator in terms of trajectories with fixed start and end points, $s'$ and $s$, as follows
\begin{equation}\label{endpt_approx}
\begin{array}{l}
{B_{m+1+N(l-1),\beta+1+N(\alpha-1)}=}\vspace{2mm}\\
{\displaystyle\frac{(2m+1)}{4}\int_{\Gamma}\int_{\Gamma} w(Y^{s})\tilde{P}_{m}(p_s(s,s'))b_{l}(s)\tilde{P}_{\beta}(p'_{s}(s,s'))b_{\alpha}(s')\left|\frac{\partial p'_{s}}{\partial s}\right|ds'ds.}
\end{array}
\end{equation}
The resulting boundary integral formulation containing a pair of integrals over boundary coordinates bears a resemblance to standard variational Galerkin boundary integral formulations such as in \cite{GS03}.

\subsection{Subsystems and links to SEA}

Recall the splitting into subsystems $\Omega_{i}$, $i=1,..,N_{\Omega}$ introduced earlier. The dynamics in each subsystem are considered separately so that both variability in the wave velocity $c_{i}$ and non-convex domains may be handled easily. Coupling between sub-elements can then be treated as losses in one subsystem and source terms in another. Typical subsystem interfaces are surfaces of reflection/ transmission due to sudden changes in material parameters or local boundary conditions. We describe the full dynamics in terms of subsystem boundary operators $\mathcal{B}_{ij}$; flow between $\Omega_{j}$ and $\Omega_{i}$ is only possible if $\Omega_{i}\cap\Omega_{j}\neq\emptyset$ and one obtains an integral kernel for $\mathcal{B}_{ij}$ of the form
\begin{equation}\label{fpo-ij}
K_{ij}(X_{i}^{s},X_{j}^{s})=w_{ij}(X_{i}^{s})\delta(X_{i}^{s}-\phi^{\omega}_{ij}(X_{j}^{s})),
\end{equation}
where $\phi_{ij}^{\omega}$ is the boundary map in subsystem $j$ mapped onto the boundary of the adjacent subsystem $i$ and $X_{i}^{s}$ are the boundary coordinates of $\Omega_{i}$. The weight $w_{ij}$ contains reflection and transmission coefficients characterizing the coupling at the interface between $\Omega_{j}$ and $\Omega_{i}$. It also contains a damping factor of the form $\exp(-\mu_{i} L)$ where $\mu_{i}$ is the damping coefficient in $\Omega_{i}$ as before and $L$ is the length of the trajectory from $s'$ to $s$.

Repeating the steps in the previous subsection but instead using the operator above results in a basis function representation, see \cite{GT09, DC11} for details of this process. Here we employ a boundary mesh ensuring that the boundary of an interface between two subsystems only intersects element boundaries and not their interiors. An SEA treatment emerges when approximating the individual operators $\mathcal{B}_{ij}$ in terms of constant functions only \cite{GT09}. Here this corresponds to an approximation in terms of the lowest order basis functions only, together with a coarse spatial mesh with only one element per subsystem, or more typically a piecewise constant approximation on a mesh with continuity enforced within each subsystem. In this case the matrix $B$ can be reduced to one element per subsystem interaction, say between subsystem $i$ and subsystem $j$, possibly with $i=j$. The matrix entry gives the mean transmission rate from subsystem $j$ to subsystem $i$. It is thus equivalent to the coupling loss factor used in standard SEA equations \cite{RL95}. The resulting full $N_{\Omega}$-dimensional $B$ matrix yields a set of SEA equations using the relation (\ref{Neum-ser}) after mapping the boundary densities back into the interior.

\section{\label{sec:numerics} Numerical results}
\subsection{Verification in 2D for Coupled Two-Cavity Problems}

In this section we consider two-dimensional polygonal domains whose boundaries are meshed by subdividing each side into equidistant sections governed by a mesh parameter $\Delta x$. The number of elements on any given side is computed using the integer part of the side length divided by $\Delta x$. The Jacobian from (\ref{endpt_approx}) is written in the form
\begin{equation}\label{2DJac}
\left|\frac{\partial p'_{s}}{\partial s}\right|=\frac{k_{j}(n\cdot(r-r'))(n'\cdot(r-r'))}{L^3},
\end{equation}
where $n$ and $n'$ are the internal unit normal vectors to $\Gamma$ at $r$ and $r'$, respectively.  In order to treat the corner singularities in (\ref{2DJac}), Gaussian quadrature is employed where end-points are not included as quadrature points. The convergence of the quadrature rules is still slow due to the peak in the integrand at corners. Telles' transformation techniques are employed to speed up the convergence \cite{JT87}.

\begin{figure}
\centering
\includegraphics[width=10cm]{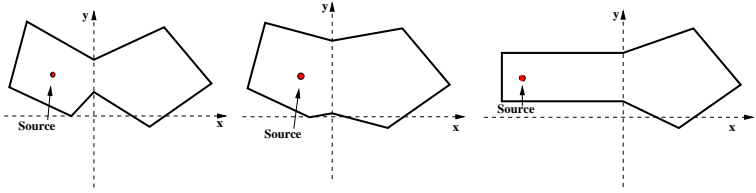}
\caption{Coupled two-domain systems: configurations A, B and C respectively.}
\label{twodomplots}
\end{figure}

A number of two-cavity systems are considered as shown in Fig.\ \ref{twodomplots} with Dirichlet boundary conditions on the outer boundary. Configuration A features irregular shaped, well separated pentagonal subsystems. In configuration B the size of the interface between the subsystems is increased reducing their dynamical separation. Configuration C includes a rectangular left-hand subsystem channelling rays out of the subsystem and introducing long-range correlations in the dynamics. In addition, the source is further from the intersection of the two subsystems. Note that SEA results are in general insensitive to the position of the source, whereas actual trajectory calculations may well depend on the exact position.

In \cite{GT09, DC11} it is demonstrated, as expected, that SEA works well for configuration A, but not so well for configurations B and C. In this communication we seek to verify our new approach against results from previous work. In particular we discuss the relative computational efficiency of the new and old approaches and how they scale as the level of precision in the model is increased. Energy distributions have been studied as a function of the frequency with a hysteretic damping factor $\eta=0.01$, where $\mu_{i}=\omega\eta/(2c_{i})$ for $i=1,2$. Here and in the remainder of this section the subsystems are numbered $1,2$ from left to right. The other parameters are set to unity for simplicity, that is $\varrho_{i}=c_{i}=1$ for $i=1,2$. For this reason the subsystem interface reflection and transmission coefficients appearing in the weight term in (\ref{fpo-ij}) are simply 0 and 1, respectively.

\begin{figure}
\centering
\includegraphics[width=10cm]{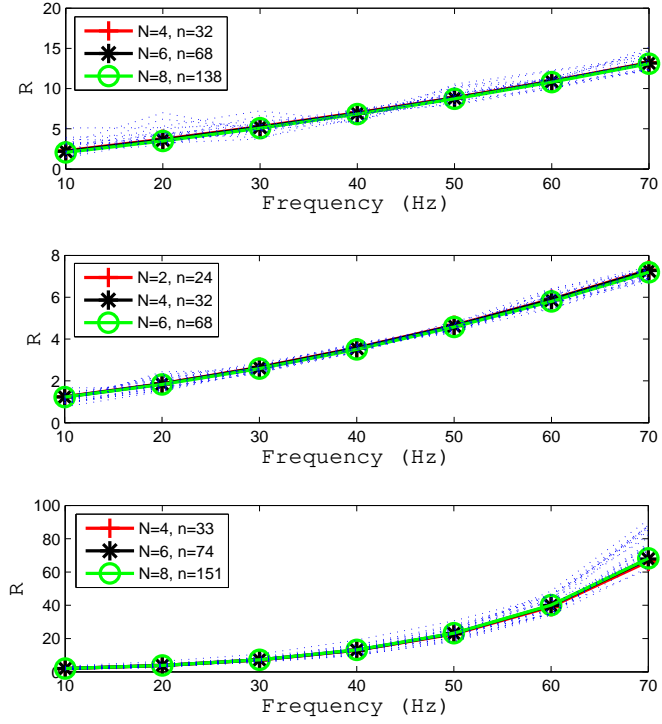}
\caption{(Color online) Ratio of total energies $\textsf{R}=\|G_{1}\|^{2}/\|G_{2}\|^{2}$ in configurations A, B and C respectively. The dotted lines correspond to FEM calculations of the full wave problem.}
\label{conAerr}
\end{figure}

Fig.\ \ref{conAerr} shows the the boundary element DEA results for configurations A, B and C compared with discontinuous Galerkin finite element method simulations as detailed in \cite{DC11}. Explicitly we compute the energy ratios between the two subsystems $R=\|G_{1}\|^{2}/\|G_{2}\|^{2}$ where
\begin{equation}\label{ennorm}
\|G_{i}\|^{2}:=\int_{\Omega_{i}}|G(r,r_{0};\omega)|^{2}dr,\hspace{2mm}i=1,2.
\end{equation}
The dotted lines each represent a simulation at a different frequency in the range $\pm 5$Hz of the frequencies used for the boundary element DEA calculations. In all three cases good convergence of the method is demonstrated. For configuration B, shown in the central subplot, the results converge with a slightly lower order of approximation. This may be due to the irregular geometry and the wide opening linking the subsystems meaning that the energy is very evenly distributed throughout the whole domain. Hence lower order spatial approximations will be reasonably good.

\begin{table}\caption{Computational times for Configuration A with $f=10$Hz}\label{con1comptimes} 
\centering
 \begin{tabular}{cc|cc}
  {n} & {N} & \multicolumn{2}{c}{Total Computational times (s):}\\
  {} & {} & {Boundary element} & {Chebyshev}\\
  \hline
  {32} & {4} & {140} & {590} \\
  {68}  & {6} & {280} &   {3910}\\
  {138}  & {8} & {850} &   {31290}\\
 \end{tabular}
 \end{table}

Table \ref{con1comptimes} shows the total computational times for the $10$Hz calculation in Fig.\ \ref{conAerr} using both boundary element DEA and comparing with a previous approach where a Chebyshev basis is employed in full phase space \cite{DC11}. The computations were performed using a desktop PC with a 2.83 GHz dual core processor, although the code was not parallelized. The total computational expense is considerably reduced using the current boundary element DEA approach. In addition the computational cost of boundary element DEA is growing more slowly as the precision of the model is increased. This will be very important for the three-dimensional case where the number of degrees of freedom in the model will increase more quickly. When we consider that the Chebyshev algorithm was already a considerable saving on the original DEA methods discussed in \cite{GT09} one can see how far we have come. Since the parameters for configurations B and C are similar to those for configuration A, the computational times are roughly the same for the same orders of approximation.

\subsection{Applications in 3D}

In this section we consider some three-dimensional domains whose boundaries have been triangulated using the Tetgen freeware automated mesh generating package (\verb"http:\\tetgen.berlios.de"). The Jacobian from (\ref{endpt_approx}) may be computed using standard formulae for global, local and polar coordinate transformations along with several applications of the chain rule. As before the Jacobian introduces singularities in the integrals along edges and at vertices of the domain. Again Gaussian quadrature and a carefully chosen application of Telles' transformation techniques are employed to ensure relatively fast convergence of the numerical integration procedures.

\begin{figure}
\centering
\includegraphics[width=12cm]{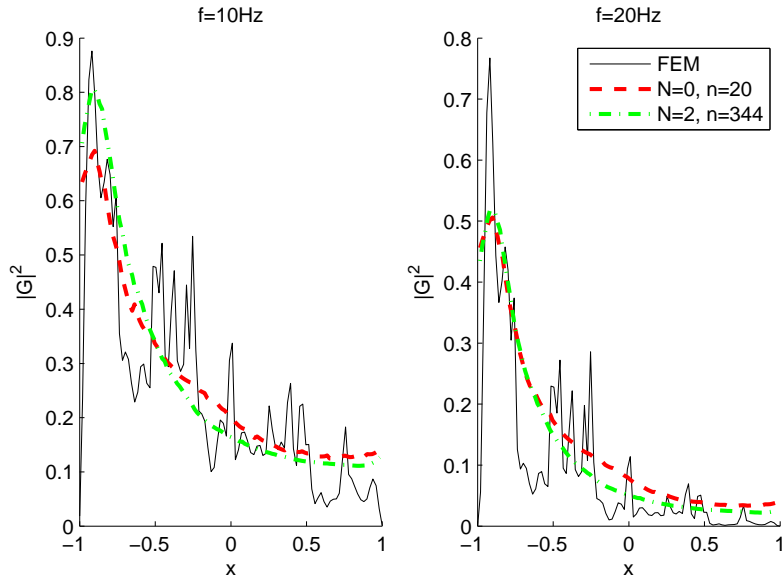}
\caption{Energy distribution along the x-axis in the cuboid example.}
\label{cuboid}
\end{figure}

The first example we consider is that of a cuboid $(x,y,z)\in(-1,1)\times(-0.5,0.5)\times(-0.5,0.5)$ with Dirichlet boundary conditions. The source point is taken as $(-0.9, 0.1, 0.1)$ and the same frequency and damping correspondence is used as in the two-dimensional examples and $\varrho=c=1$. Figure \ref{cuboid} shows the computed energy distributions inside the cuboid along the x-axis. The method is compared against discontinuous Galerkin finite element method (FEM) computations, which are averaged over $17$ frequencies within $\pm2$Hz of the (central) frequency used for the boundary element DEA computation. Further details of the FEM techniques employed here can be found in \cite{DC11} and references therein. The dashed line shows an approximation with a coarse mesh and where the energy density is assumed constant over all possible directions of rays approaching the boundary from the interior. The computation time for such an approximation is typically around a minute per frequency. The dash-dot line shows a higher order approximation where we have refined the mesh until the solution appears reasonably converged by eye, in this case 344 elements were employed. Also due to the relatively low dissipation levels in these plots a low order approximation in momentum (quadratic) was sufficient to give reasonably converged results. The computation time for this plot was approximately 16 hours using a parallel machine with two quad-core processors.

It has been demonstrated that in the low damping regime SEA can provide reasonably good approximations even in regular structures \cite{DC11}, which explains why the coarse approximation here is still reasonably good. However, one notices an improvement in the match with the FEM data at both the peak and tail of the plot for both frequencies considered when the higher order approximation method is employed. There is, however, a significant computational cost associated with this increased precision.
\begin{figure}
\centering
\includegraphics[width=12cm]{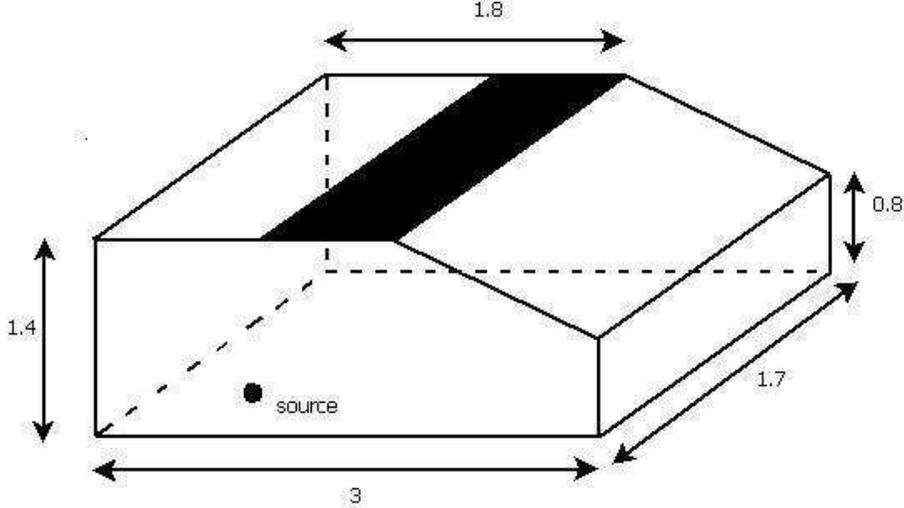}
\caption{The car cavity showing the source point and the open roof (black) acting as absorbing boundary.}
\label{carcavd}
\end{figure}

The second example we consider is an open car cavity as discussed in \cite{JR10} and shown in Fig.\ \ref{carcavd}. The source point is located on the base of the cavity at $(0.6,0.0,0.4)$. Again we consider Dirichlet boundary conditions except along the roof, which is assumed to be non-reflecting along the subsection shown in black in Fig.\  \ref{carcavd}, between $x=1.0$ and $x=1.8$. Physically this corresponds to an opening in the cavity with identical media in both the interior and exterior.

\begin{figure}
\centering
\includegraphics[width=12cm]{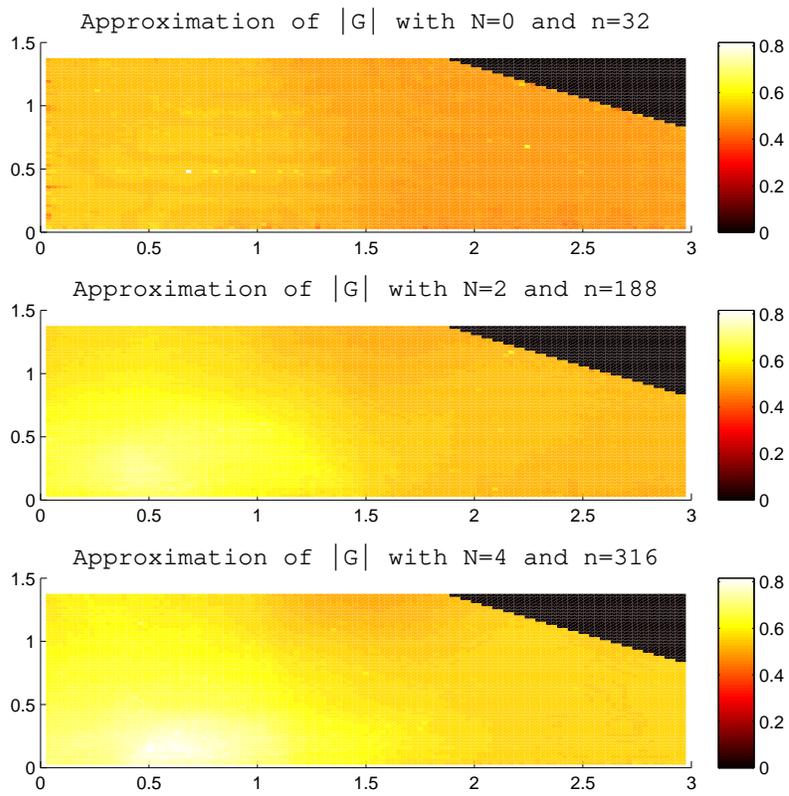}
\caption{Amplitude $|G|$ along the plane z=0.85 for the car cavity example.}
\label{carcav}
\end{figure}

Figu.\ \ref{carcav} shows the amplitude $|G|$ plotted in the interior of the cavity along the midplane $z=0.85$. The jagged diagonal edge shown in Fig.\ \ref{carcav} is merely an artifact of the plotting and interpolation of the amplitude at discrete points, and is actually smooth in the model as shown in Fig.\ \ref{carcavd}. No damping is incorporated in this example and energy losses only occur through the open roof, meaning that the plots here are now wavenumber independent. The three subplots show successively higher order approximations from upper to lower and convergence in the plots is evident due to the increased similarity between the lower two plots. Directivity in the wave field plays a much stronger role in this example due to the localised dissipation at the opening of the cavity. For this reason it was necessary to employ a higher order momentum basis approximation than before with $N=4$ before the plot appears reasonably converged. The computational times were approximately one minute for the upper subplot, eleven hours for the central subplot and three days for the lower plot, again using a parallel machine with two quad-core processors.

The upper plot in Figure \ref{carcav} has a markedly different appearance to the other subplots showing that the most coarse approximation is not good in this case. One reason for this is the much stronger directivity of the wave field compared with the previous example. One can clearly see how in the upper plot the solution is more slowly varying and distributed more uniformly as you move away from the source point. In the lower plot one sees a noticeable dip in the amplitude close to the non-reflecting boundary. Also the increased intensity around the source point stretches more in the horizontal direction than the vertical direction in the lower plot, but is more evenly distributed in the other plots. In all three subplots the wave amplitude is greater in the region to the left of the opening (for $x<1$) since in this region there are many possible ray trajectories that remain trapped in the cavity for a long time before exiting through the opening. This is also true for the region $y<0.8$ and hence the greatest intensities are observed in the intersection of these two regions. In billiard dynamics these trajectories are known as near-bouncing-ball orbits, see for example \cite{GT97}.


\section{Conclusions}
A new approach to determining the distribution of mechanical and acoustic wave energy in complex built up structures has been discussed. The methodology has been carefully chosen to permit application to two or three dimensional problems. Using boundary element meshes for three dimensional problems renders the method applicable to general domains and removes the need to determine an orthogonal spatial basis for each geometrically different example. The application of the method to some well studied two-dimensional examples has shown it to be relatively efficient and scale favourably as the number of degrees of freedom in the model is increased compared with previous DEA approaches. Examples in three dimensions were also considered showing both the applicability and versatility of the method, but also its high computational cost as the number of degrees of freedom is increased. We have also seen however that in some cases a low order and fast computation yields reasonably good results. The suitability of the method for parallel processing means that with greater computing resources it has the potential to be employed in larger and more complicated configurations than those considered here.

\section*{Acknowledgement}
\noindent Support from the EPSRC (grant EP/F069391/1) and the EU (FP7IAPP grant MIDEA) is gratefully acknowledged.
The authors also wish to thank inuTech Gmbh, N\"{u}rnberg for Diffpack guidance and licences
and Hanya Ben Hamdin and Dmitrii Maksimov for helpful discussions.

\end{document}